\documentclass[12pt,]{article}
 \usepackage{graphicx}
 \usepackage[cp1251]{inputenc}  
\begin{document}

\begin{center}{\Large\bf Relationship between Poincar\'e Sections and Spectral Characteristics
of Orbits of Globular Clusters in the Central Region of the Galaxy}
\end{center}
  \bigskip
\centerline{\bf
            A. T. Bajkova\footnote [1]{e-mail: bajkova@gaoran.ru},
            A. A. Smirnov,
            V. V. Bobylev
            }
 \bigskip
 \centerline {\small \it Central Astronomical Observatory, Russian Academy of Sciences, Pulkovo, 196140 Russia}
 \bigskip

In the paper, orbital dynamics, regular or chaotic, of globular clusters (GCs) in the central region
of the Galaxy, which is subject to the greatest influence of the rotating bar, has been studied. Such methods
for determining chaos as Poincar\'e sections and spectral methods have been compared. The relationship
between the Poincar\'e sections and the spectral characteristics of the orbits has been estimated. The sample
includes 45 globular clusters in the central region of the Galaxy with a radius of 3.5 kpc. To form the 6D-phase
space required for integrating the orbits, the most accurate astrometric data to date from the Gaia satellite, as
well as new refined average distances, have been used. The following, most realistic, bar parameters have been
adopted: mass $10^{10} M_\odot$, length of the
major semi-axis of the bar model in the form of a triaxial ellipsoid is 5 kpc, angle of
rotation of the bar axis is $25^o$, rotation velocity is 40 km s$^{-1}$ kpc $^{-1}$.
The result of the study is that a 100\% correlation between the classification by Poincar\'e sections and the spectral characteristics
of the orbits has been established. Consequently, the classification by Poincar\'e sections can be
replaced by a more visual analysis of the amplitude spectra of the orbits. Thus, two lists of GCs: with regular
and chaotic dynamics have been compiled. The GCs with varying degrees of orbital chaos have separately
been distinguish.

\bigskip\noindent


DOI: 10.1134/S1063772925702324

\newpage

\section*{INTRODUCTION}

This study is essentially a continuation of [1–3] devoted to the study of the orbital dynamics (regular or
chaotic) of globular clusters in the central region of the Galaxy. As in previous studies, the sample includes 45
globular clusters in the central region of the Galaxy with a radius of 3.5 kpc. To form the 6D-phase space
required for orbit integration, the most accurate astrometric data to date from the Gaia satellite [4], as well
as new refined mean distances [5], were used. The following most realistic parameters of the bar that are
known from the literature [6, 7] are adopted: the mass is $10^{10} M_\odot$, the length of the major semi-axis is 5 kpc,
the rotation angle of the bar axis is $25^o$, and the rotation speed is 40 km s$^{-1}$ kpc $^{-1}$.

Since GCs in the central region of the Galaxy are subject to the greatest influence from the elongated
rotating bar, the question of the nature of the orbital motion of GCs (regular or chaotic) is of great interest.
For example, in [8], it is shown that the main share of chaotic orbits must be precisely in the bar region.

This study is aimed at establishing the connection between Poincar\'e sections and spectral characteristics
of orbits as functions of time. Spectral methods include, in particular, the frequency method [9–13].
The authors of these studies showed that it is possible to measure the stochasticity of the orbit based on the
shift of fundamental frequencies determined over two consecutive time intervals. Another method of this
class is our recently proposed method [3] based on calculating the orbital power spectrum as a function of
time and calculating the entropy of the power spectrum as a measure of orbital chaos.

The paper is structured as follows. In the first section, the accepted potential models: the axisymmetric
potential and the non-axisymmetric potential including a bar is briefly described. In the second section,
links to the used astrometric data, as well as the method for forming the GC sample, are provided. In
the third section, the methods used to estimate the regularity/chaotic nature of motion: the Poincar\'e section
method, the frequency method, and the spectral methods we proposed are described. In the fourth section,
the obtained results are analyzed and a connection between the Poincar\'e cross sections and the spectral
characteristics of the orbits are established. The main results of the study are formulated in the Conclusions
section.

\section{GALACTIC POTENTIAL MODEL}

\subsection{Axisymmetric Potential}

The axisymmetric gravitational potential of the
Galaxy traditionally used by us (see, for example, [2])
for integrating the orbits of GCs is represented as the
sum of three components: the central spherical bulge $\Phi_b(r(R,Z))$, disk $\Phi_d(r(R,Z))$, and a massive spherical
dark matter halo $\Phi_h(r(R,Z))$:

 \begin{equation}
 \begin{array}{lll}
  \Phi(R,Z)=\Phi_b(r(R,Z))+\Phi_d(r(R,Z))+\Phi_h(r(R,Z)).
 \label{pot}
 \end{array}
 \end{equation}

 A cylindrical coordinate system ($R,\psi,Z$) with the origin
of coordinates at the center of the Galaxy is used
here. In a rectangular coordinate system $(X,Y,Z)$ with
the origin at the center of the Galaxy, the distance to
the star (spherical radius) will be equal to
$r^2=X^2+Y^2+Z^2=R^2+Z^2$, while the $X$ axis is directed from
the Sun to the galactic center, the $Y$ axis is perpendicular
to the axis in the direction of rotation of the
Galaxy, and the $Z$ axis is perpendicular to the galactic
plane $(X,Y)$ towards the north galactic pole. The gravitational
potential is expressed in units of 100 km$^{2}$ s$^{-2}$,
distances are in kpc, masses are in units of galactic
mass, $M_{gal}=2.325\times 10^7 M_\odot$, corresponding to the
gravitational constant of $G=1$.

Axisymmetric bulge potentials $\Phi_b(r(R,Z))$ and
disk $\Phi_d(r(R,Z))$ are presented in the form proposed
in [14]:
 \begin{equation}
  \Phi_b(r)=-\frac{M_b}{(r^2+b_b^2)^{1/2}},
  \label{bulge}
 \end{equation}
 \begin{equation}
 \Phi_d(R,Z)=-\frac{M_d}{\Biggl[R^2+\Bigl(a_d+\sqrt{Z^2+b_d^2}\Bigr)^2\Biggr]^{1/2}},
 \label{disk}
\end{equation}
where $M_b$ and $M_d$ are the masses of components; and $b_b$, $a_d$, and $b_d$ are the scale parameters of components
in units of kpc. The halo component (NFW) is presented
according to [15]:
 \begin{equation}
  \Phi_h(r)=-\frac{M_h}{r} \ln {\Biggl(1+\frac{r}{a_h}\Biggr)},
 \label{halo-III}
 \end{equation}
where $M_h$ is the weight, and $a_h$ is the scale parameter.
In Table 1, the values of the parameters of the adopted
model of the galactic potential are shown.

\begin{figure*}
{\begin{center}

\includegraphics[width=0.4\textwidth,angle=-90]{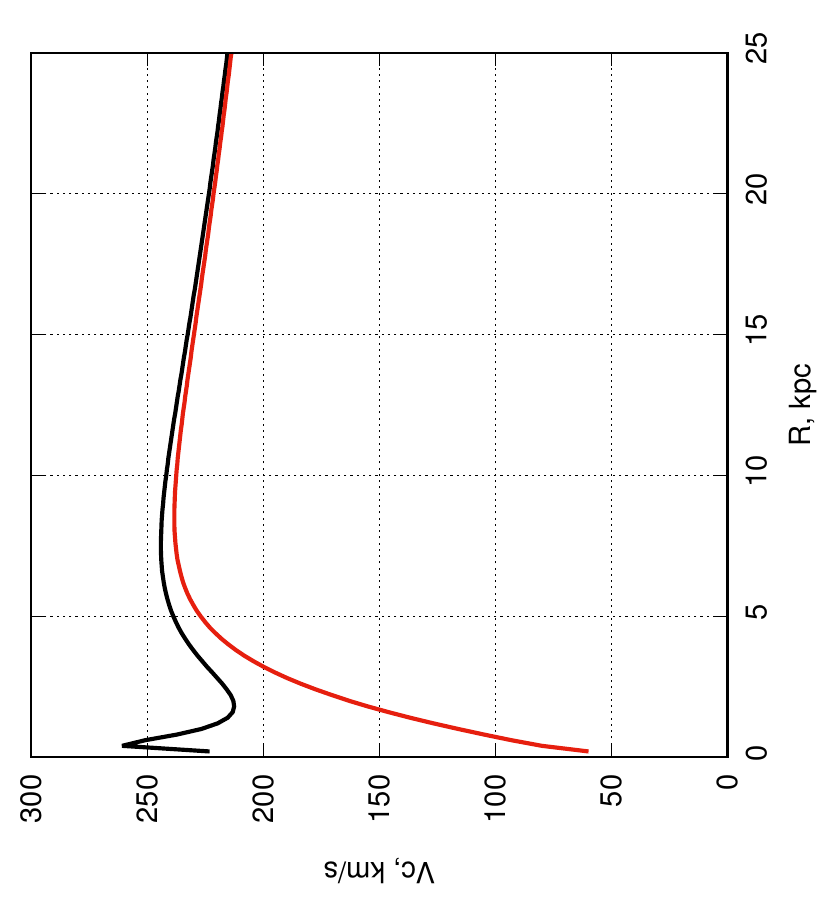}

\bigskip

\caption{\small Rotation curve of the Galaxy with an axisymmetric potential without a bar (black line) and a non-axisymmetric potential
including a bar (red line).}
\label{fcomp}
\end{center}}
\end{figure*}

\subsection{Bar Model}

The triaxial ellipsoid model was chosen as the central
bar potential [6]:
\begin{equation}
  \Phi_{bar} = -\frac{M_{bar}}{(q_b^2+X^2+[Ya/b]^2+[Za/c]^2)^{1/2}},
\label{bar}
\end{equation}
where $X=R\cos\vartheta$, $Y=R\sin\vartheta$, $a, b, c$ are three
semi-axles of the bar, $q_b$ is the scale parameter of the
bar (length of the largest semi-axis of the bar);
$\vartheta=\theta-\Omega_{b}t-\theta_{b}$, $tg(\theta)=Y/X$,
$\Omega_{b}$ is the circular velocity
of the bar, $t$ is the integration time, and $\theta_{b}$ is the orientation
angle of the bar relative to the galactic axes $X,Y$
that is measured from the line connecting the
Sun and the center of the Galaxy (axis $X$) to the major
axis of the bar in the direction of rotation of the
Galaxy.

Based on information in the literature, in particular
in [7], the following were used as bar parameters:
$M_{bar}=430 M_{gal}$, $\Omega_{b}=40$ km s$^{-1}$ kpc $^{-1}$, $q_b=5$ kpc,
and $\theta_{b}=25^o$. The accepted bar parameters are listed
in Table~1.

 {\begin{table}[t]                                    
 \caption[]
 {\small\baselineskip=1.0ex
Values of the parameters of the galactic potential
model, $M_{gal}=2.325\times 10^7 M_\odot$
  }
 \label{t:model-III}
 \begin{center}\begin{tabular}{|c|r|}\hline
 $M_b$ &   443 M$_{gal}$ \\
 $M_d$ &  2798 M$_{gal}$ \\
 $M_h$ & 12474 M$_{gal}$ \\
 $b_b$ & 0.2672 kpc  \\
 $a_d$ &   4.40 kpc  \\
 $b_d$ & 0.3084 kpc  \\
 $a_h$ &    7.7 kpc  \\
\hline\hline
 $M_{bar}$ & 430 M$_{gal}$ \\
 $\Omega_b$ & 40 km s$^{-1}$ kpc $^{-1}$ \\
 $q_b$     &  5.0 kpc  \\
 $\theta_{b}$ &  $25^o$   \\\hline
 $a/b$ & 2.38  \\
 $a/c$ & 3.03  \\
    \hline
 \end{tabular}\end{center}\end{table}}

To integrate the equations of motion, we used
Runge–Kutta algorithm of the fourth order.

The value of the peculiar velocity of the Sun relative
to the local standard of rest was taken to be equal
to $(u_\odot,v_\odot,w_\odot)=(11.1,12.2,7.3)\pm(0.7,0.5,0.4)$ km s$^{-1}$
according to [16]. The elevation of the Sun above the
plane of the Galaxy is taken to be 16 pc in accordance
with [17].

For comparison, the obtained model rotation
curves: an axisymmetric potential (black line) and a
potential with a bar (red line) are shown in Fig.~1.

\section{DATA}

The data on the proper motions of GCs are taken
from a new catalogue [4] compiled based on observations
of Gaia EDR3. The average values of distances to
globular clusters are taken from [5].

The GCs catalogue [18] at our disposal contains
152 objects. Globular clusters from this set belonging
to the bulge/bar region were selected in accordance
with a purely geometric criterion considered in [19]
and also used by us in [20]. It is very simple and consists
of selecting GCs whose apocentric distance of
orbits does not exceed the bulge radius, which is usually
taken to be 3.5 kpc. Orbits were calculated in an
axisymmetric potential. The full list of 45 objects in
our sample is presented in Table 2, which the results of
the analysis of the orbital chaoticity/regularity of
GCs (the first column gives the ordinal number of the
GCs, the second column gives the name of the GCs)
are shown.

\section{METHODS}

We recall the main provisions of the methods considered
in this study for determining the nature of
orbital dynamics: chaotic or regular. A more detailed
description is given in [1–3].

\subsection{Poincar\'e Sections}

The algorithm used to construct the mappings is as
follows [21]:

1. We consider the phase space $(X,Y,V_x,V_y)$.

2. We exclude $V_y$, using the law of conservation of
the generalized energy integral (Jacobi integral) and
move into space $(X,Y,V_x)$.

3. We define a plane of $Y=0$, we will designate
the points of intersection with the orbit on the plane
$(X,V_x)$. We take only those points, where $V_y>0$.

If the intersection points of the plane form a continuous
smooth line (or several separated lines), then
the motion is considered regular. In the case of chaotic
motion, instead of being located on a smooth curve,
the points fill a two-dimensional region of phase
space, and sometimes the effect of points sticking to
the boundaries of islands corresponding to ordered
motion occurs [22].

In this paper, we present the Poincar\'e sections
obtained by us in [1].

\subsection{Frequency Method}

The method consists of measuring the orbital
chaos based on the shift of fundamental frequencies
determined over two consecutive time intervals. For
each frequency component $f_i$, a parameter called frequency
drift is calculated:
\begin{equation}
\label{freq}
\lg(\Delta f_i)=\lg|\frac{\Omega_i(t_1)-\Omega_i(t_2)}{\Omega_i(t_1)}|,
\end{equation}
where $i$ defines the frequency component in Cartesian
coordinates (i.e. $\lg(\Delta f_x), \lg(\Delta f_y)$, and $\lg(\Delta f_z)$)).
Then, the largest value of these three frequency drift
parameters is attributed to the frequency drift
parameter $\lg(\Delta f)$. The higher the value $\lg(\Delta f)$, the
more chaotic the orbit. In order to achieve high accuracy,
we took an integration time of 120 billion years,
almost an order of magnitude greater than the age of
the Universe. In this study, we also used the classification
results given in [1].

\subsection{Spectral Analysis of Orbits}

The spectral analysis of orbits proposed by us in [3]
is based on the calculation of the modulus of the discrete Fourier transform (DFT) of uniform time series
of radial distances of orbital points from the center of
the Galaxy, $r_n$, calculated based on their $X, Y, Z$
galactic coordinates,$X(t_n), Y(t_n), Z(t_n)$ as functions
of time: $r(t_n)=\sqrt{X(t_n)^2+Y(t_n)^2+Z(t_n)^2}$, where $n=0,...,N-1$
($N$ is the length of the row).

Thus, the formula for the DFT modulus (amplitude
spectrum) of a sequence $r_n$ will look like this:
\begin{equation}
\label{Equ1}
\bar{r}_k=|\frac{1}{N} \sum_{n=0}^{N-1} r(t_n)\exp{(-\jmath\frac{2\pi\times n\times k}{N})}|,~~~ k=0,..., N-1.
\end{equation}

In this case, the length of the row is selected equal
to $N=2^\alpha$, where $\alpha$ is the positive integer ($>0$ ), so
that the fast Fourier transform algorithm can be used
to calculate the DFT. The required length of the series
is achieved by supplementing the real series with zeros.
In our case, the length of the actual sequences is
120 000, since we integrate the orbits back 120 billion years with an integration interval of 1 million years.
Before calculating the DFT, we first center the coordinate
series (i.e., get rid of the constant component),
then complement the resulting sequence $r_n$ zero readings
at $n > 120000$ until the length of the entire analyzed
sequence is reached, $N = 262144 = 2^{18}$.
Note that supplementing the initial sequence with zeros is
also useful from the point of view of increasing the
accuracy of the coordinates of the spectral components.
Since the interval between the readings of the
sequences in time is equal to $\delta t = 0.001$ billion years,
then the analyzed frequency range, which is a periodic
function, is $F=1/\Delta_t=1000$ Gyr$^{-1}$. The frequency
discrepancy is $\Delta_F= F/N \approx 0.03815$ Gyr$^{-1}$. Next, for
convenience, we will indicate on the graphs not the
physical frequencies, but the sample numbers $k$ (or $K$)
of the discrete Fourier transform (2). Transition from $k$
to the physical frequency can be produced by the
formula: $f = k \times \Delta F \approx k \times 0.003815$. Next, the obtained
power spectrum of the GCs orbit is normalized so that
the maximal value is equal to unity.

The decision on the nature of the orbital dynamics
of GCs is determined by calculating the Shannon
entropy of the normalized amplitude spectrum $\bar{r}_k$ as
measures of chaos [23]:
\begin{equation}
\label{Equ2}
E_r=-\frac{1}{M}\sum_{k=0}^{N-1} \bar{r}_k \ln(\bar{r}_k),
\end{equation}
where $M$ is a scale factor that is introduced for the
convenience of presenting numerical results.

Obviously, the higher the entropy value, the higher
the degree of chaos of the orbit.

In this case, we analyze both the reference orbits
and the shadow ones obtained by perturbing the initial
phase point, as accepted in [1–3], as follows:: $X_1 = X_0+X_0\times 0.00001, Y_1 = Y_0+Y_0\times 0.00001, and Z_1 = Z_0+Z_0\times 0.00001$.

\subsection{Spectral Analysis of Orbital Coordinates $X$ and $V_x$}

In this work, with the aim of establishing a connection
between the Poincar\'e sections on the plane $(X,V_x)$
and the spectral characteristics of the orbits,
we propose to calculate the modulus of the discrete
Fourier transform of uniform time series of coordinates $X(t_n)$
and $V_x(t_n)$:
\begin{equation}
\label{Equ1}
\bar{X}_k=|\frac{1}{N} \sum_{n=0}^{N-1} X(t_n)\exp{(-\jmath\frac{2\pi\times n\times k}{N})}|,~~~ k=0,..., N-1.
\end{equation}

\begin{equation}
\label{Equ1}
\bar{V_x}_k=|\frac{1}{N} \sum_{n=0}^{N-1} V_x(t_n)\exp{(-\jmath\frac{2\pi\times n\times k}{N})}|,~~~ k=0,..., N-1.
\end{equation}

As an example, the obtained amplitude spectra for
two GCs: NGC 6266 and NGC 6355 with regular and
chaotic dynamics, respectively, are shown in Fig.~2. As
can be seen from the figure and as the analysis of the
GC spectra of the entire sample shows, the spectra of
the coordinates $X$ and $V_x$ are similar. Therefore, in
order to save space, we present below only the spectra
of -coordinates in Fig.~3.

The same as in the case of the spectral method proposed
in [3], regular orbits correspond to narrow linear
spectra, while chaotic orbits correspond to wide spectra.
As will be shown in the next section, this follows,
from a comparison of the Poincar\'e sections with the
results of a spectral analysis, which is the goal of this
paper.

\begin{figure*}
{\begin{center}

     \includegraphics[width=1.0\textwidth,angle=0]{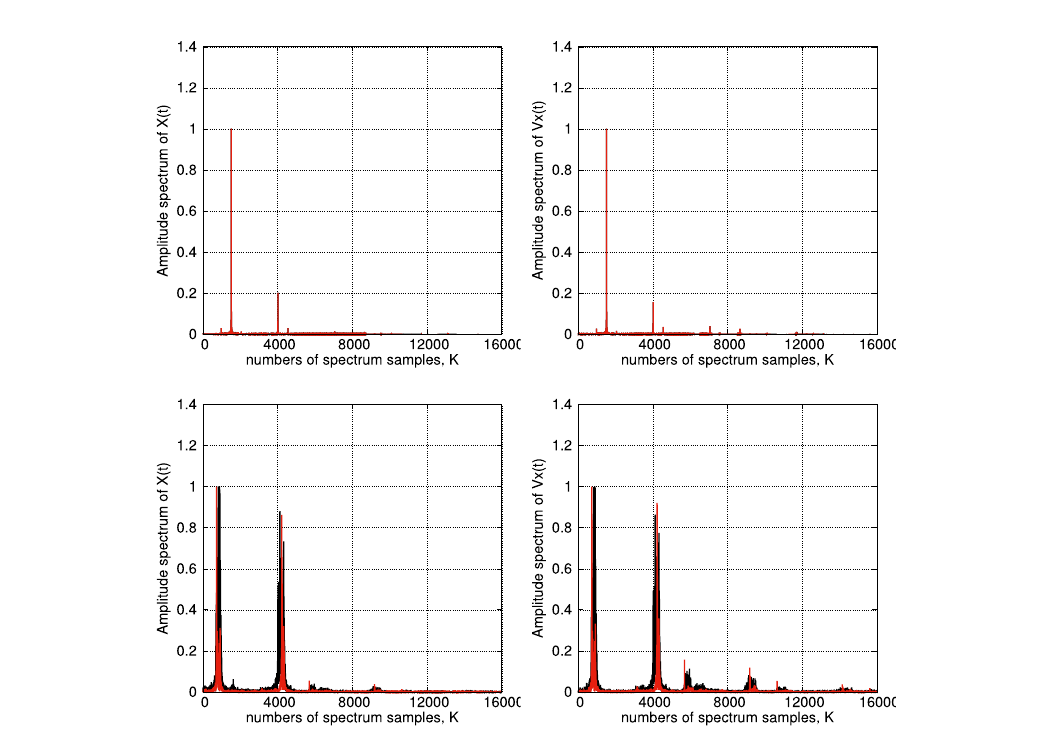}

\bigskip
\caption{\small Normalized power spectra of -coordinates (left), -coordinates (right) of the reference and shadow orbits as functions
of time that are shown in black and red, respectively. The upper panels refer to GC of NGC 6266 with regular dynamics, and the
lower panels refer to GC of NGC 6355 with chaotic dynamics.}
 \label{f1}
\end{center}}
\end{figure*}

\section{RESULTS}

A graphical representation of the spectral analysis
results in comparison with the results obtained previously
[1–3] (Poincar\'e sections, frequency method,
and visual analysis) for the entire sample of 45 GCs is
given in Fig.~3, and the results of the classification of
orbits into regular (R) and chaotic (C) are given in
Table~2 (the first column, as a reminder, shows the
serial number of the GCs, while the second column
shows the name of the GCs).

The proposed method was applied to both reference
orbits and shadow orbits. The integration of
orbits was carried out, as was already noted above,
back 120 billion years.

In Fig.~3, it is shown from left to right: (1) projections
of orbits onto the plane ($X-Y$), (2) radial values
of the initial (reference) and perturbed (shadow) orbits
depending on time (reference orbits are shown in yellow,
shadow orbits are in purple), (3) Poincar\'e sections
on the plane ($X,V_x$), (4)$X$ - coordinates of Poincar\'e
sections, (5)$V_x$ - coordinates of Poincar\'e sections,
(6) normalized power spectra of $X$ - values of the reference
and shadow orbits as functions of time that are
shown in black and red, respectively, and (7) illustration
of the frequency method (the power spectrum of
the first half of the time series is shown in red and the
second half is shown in black). The names of the GCs
are shown in the second panels from the left.

The most obvious illustration of the discrepancy
between the reference and shadow phase points are
columns 1 and 2 in Fig.~3, which list the reference and
shadow orbits for each GC in the order (top to bottom)
as they are listed in Table~2. The first column contains
$X, Y$ projections of orbits constructed in the rotating
bar system over a time interval of $[-11,-12]$ billion
years. The second column shows the radial values of
the orbit $r(t)$ on the interval of $[0,-12]$ billion years
that is comparable to the age of both GCs and the Universe.
In these graphs, reference orbits are shown in
yellow, and shadow orbits are shown in purple. As can
be seen, many objects on the graphs have purple color
only. This means that the shadow orbit is almost identical
to the reference orbit (yellow lines are covered
with purple ones). Such objects include GCs with regular
orbits. In the graphs of GCs with chaotic orbits,
both purple and yellow lines are visible, which makes
it possible to judge qualitatively the degree of chaos of
the orbits.

In the third column, the Poincar\'e sections on the
plane $(X,V_x)$ are shown. Dependence of coordinates $X$
and $V_x$ on the counting number are given in the
fourth and fifth columns, respectively. The regularity
of the distribution of coordinates $X$ and $V_x$ characterizes
the regularity of orbital dynamics, and this is
reflected in the Poincar\'e sections.

As in the case of the spectral method proposed in
[3], as well as the frequency method, the normalized
amplitude spectra of the reference and shadow orbits
that are given in the sixth and seventh columns have
the character of line spectra for GCs with regular
dynamics and broad spectra for GCs with chaotic
dynamics.

The obtained amplitude spectra of coordinates $X$
and $V_x$ show a one-hundred-percent correlation with
the nature of the distribution of points on the Poincar\'e
sections, which is reflected in the sixth and seventh
columns of Table~2, which also present the results of
the classification of GCs with regular and chaotic
orbits from previous studies [1, 3], obtained using: the
spectral-entropy method [3] (third column of the
table), the frequency method (fourth column) [1], and
the visual method (fifth column) [1]. Based on an
analysis of the table, a high correlation between the
results of GC classification by different methods (not
less than 82.5\%) is shown. Due to the established connection
between Poincar\'e sections and the spectral
method for the analysis of orbital dynamics instead of
Poincar\'e sections, the decision-making process for
which is characterized by some subjectivity, a more
visual method of the spectral analysis can be used.

From the fact that the classification by Poincar\'e
sections completely coincides with the classification
by the width of the amplitude spectrum, we have
defined two lists. The first list includes 25 globular
clusters with regular dynamics (R) (NGC 6266, Terzan
4, Liller 1, NGC 6380, Terzan 1, Terzan 5, NGC
6440, Terzan 6, Terzan 9, NGC 6522, NGC 6528,
NGC 6624, NGC 6637, NGC 6717, NGC 6723, Terzan
3, Pismis 26, NGC 6569, E456–78, NGC 6540,
Djorg 2, NGC 6171, NGC 6316, NGC 6539, and
NGC 6553) and a second list of 20 globular clusters
with chaotic dynamics (C) (NGC 6144, E452–11,
NGC 6273, NGC 6293, NGC 6342, NGC 6355, Terzan
2, BH 229, NGC 6401, Pal 6, NGC 6453, NGC 6558, NGC 6626, NGC 6638, NGC 6642, NGC
6256, NGC 6304, NGC 6325, NGC 6388, and NGC
6652). Furthermore, from the second list, we consider
it appropriate to select GCs with varying degrees of
chaos using the entropy measure. For example, we
classified GCs NGC 6144, NGC 6273, NGC 6304,
NGC 6325, and NGC 6388 as GCs with weakly chaotic
dynamics. GCs E452–11, NGC 6355, Terzan 2,
BH 229, NGC 6401, Pal 6, NGC 6558, NGC 6638,
NGC 6642, and NGC 6652 exhibit strong chaos.

\section*{CONCLUSIONS}

The following main results were obtained:

1.A direct, 100\% connection has been established
between the Poincar\'e sections of regular and chaotic
orbits and the spectral characteristics of the orbits. The
wider the spectrum, the higher the entropy and the
more chaotic the orbital character is shown by the
Poincar\'e section. Thus, for the analysis of orbital
dynamics, instead of Poincar\'e sections, the classification
by which is somewhat subjective, a more visual
method of spectral analysis of orbits can be used.

2.Based on the established relationship between
the Poincar\'e sections and the spectral characteristics
of the orbits, of 45 GCs in the central region of the
Galaxy with a radius of 3.5 kpc, a list of 25 globular
clusters with regular dynamics (R) was determined:

NGC6266, Terzan4, Liller1, NGC6380, Terzan1, Terzan5, NGC6440, Terzan6, Terzan9, NGC6522, NGC6528, NGC6624, NGC6637, NGC6717, NGC6723, Terzan3, Pismi26, NGC6569, E456-78, NGC6540, Djorg2, NGC6171, NGC6316, NGC6539, and
NGC6553

and a list of 20 globular clusters
with chaotic dynamics (C):

NGC6144, E452-11, NGC6273, NGC6293, NGC6342, NGC6355, Terzan2, BH229, NGC6401, Pal6,  NGC6453,
NGC6558, NGC6626, NGC6638, NGC6642, NGC6256, NGC6304, NGC6325,  NGC6388, and NGC6652.

3.From the list of GCs with chaotic dynamics, one
can identify GCs with varying degrees of orbital chaos
based on a comparison of the entropy measure. We
classified GCs NGC6144, NGC6273, NGC6304, NGC6325, and NGC6388 as weakly chaotic. GCs
E452–11, NGC 6355, Terzan 2, BH 229, NGC 6401,
Pal 6, NGC 6558, NGC 6638, NGC 6642, and NGC
6652 show strong chaos.

\bigskip

\noindent{\bf\Large REFERENCES}

\begin{enumerate}

 \item
A. T. Bajkova, A. A. Smirnov, and V. V. Bobylev, Publ.
Pulkovo Observ. {\bf 233}, 1 (2024); arXiv: 2406.15590 [astro-ph.GA].

 \item
A. T. Bajkova, A. A. Smirnov, and V. V. Bobylev, Publ.
Pulkovo Observ. {\bf 235}, 1 (2024); arXiv: 2412.02426 [astro-ph.GA].

 \item
A. T. Bajkova, A. A. Smirnov, and V. V. Bobylev, Publ.
Pulkovo Observ. {\bf 236}, 1 (2025).

\item
E. Vasiliev and H. Baumgardt, Mon. Not. R. Astron.
Soc. {\bf 505}, 5978 (2021); arXiv: 2102.09568 [astro-ph.
GA].

\item
H. Baumgardt and E. Vasiliev. Mon. Not. R. Astron.
Soc. {\bf 505}, 5957 (2021); arXiv: 2105.09526 [astro-ph.
GA].

\item
J. Palous, B. Jungwiert, and J. Kopecky, Astron. Astrophys.
{\bf 274}, 189 (1993).

\item
J. L. Sanders, L. Smith, N. W. Evans, and P. Lucas,
Mon. Not. R. Astron. Soc. {\bf 487}, 5188 (2019); arXiv:
1903.02008 [astro-ph.GA].

\item
R. E. G. Machado and T. Manos, Mon. Not. R. Astron.
Soc. {\bf 458}, 3578 (2016); arXiv: 1603.02294 [astro-ph.
GA].

\item
N. Nieuwmunster, M. Schultheis, M. Sormani,
F. Fragkoudi, F. Nogueras-Lara, R. Schodel, and
P. McMillan, arXiv: 2403.00761 [astro-ph.GA] (2024).

\item
M. Valluri, V. P. Debattista, T. Quinn, and B. Moore,
Mon. Not. R. Astron. Soc. {\bf 403}, 525 (2010); arXiv:
0906.4784 [astro-ph.CO].

\item
J. Laskar, Celest. Mech. Dyn. Astron. {\bf 56}, 191 (1993).

\item
M. Valluri and D. Merritt, Astrophys. J. {\bf 506}, 686
(1998).

\item
E. Vasiliev, Mon. Not. R. Astron. Soc. {\bf 434}, 3174
(2013).

 \item
M. Miyamoto and R. Nagai, Publ. Astron. Soc. Jpn.
{\bf 27}, 533 (1975).

 \item
J. F. Navarro, C. S. Frenk, and S. D. M. White, Astrophys.
J. {\bf 490}, 493 (1997).

 \item
R. Sch\"onrich, J. Binney, and W. Dehnen, Mon. Not.
R. Astron. Soc. {\bf 403}, 1829 (2010).

\item
V. V. Bobylev and A. T. Bajkova, Astron. Lett. {\bf 42}, 1
(2016).

 \item
A. T. Bajkova and V. V. Bobylev, Publ. Pulkovo Observ.
{\bf 227}, 1 (2022); arXiv: 2212.00739 [astro-ph.GA].

 \item
D. Massari, H. H. Koppelman, and A. Helmi, Astron.
Astrophys. {\bf 630}, L4 (2019).

 \item
A. T. Bajkova, G. Carraro, V. I. Korchagin, N. O. Budanova,
and V. V. Bobylev, Astrophys. J. {\bf 895}, 69 (2020).

\item
C. D. Murray and S. F. Dermott, {\it Solar System Dynamics}
(Cambridge Univ. Press, Cambridge, 2012).

\item
A. Morbidelli, {\it Modern Celestial Mechanics: Aspects of
Solar System Dynamics} (Taylor and Francis, London,
2002).

\item
O. V. Chumak, {\it Entropy and Fractals in Data Analysis}
(Inst. Komp. Issled., Moscow, 2011).

\end{enumerate}

\newpage
    \voffset=-6.40truecm

\begin{figure*}
{\begin{center}
     \includegraphics[width=1.0\textwidth,angle=0]{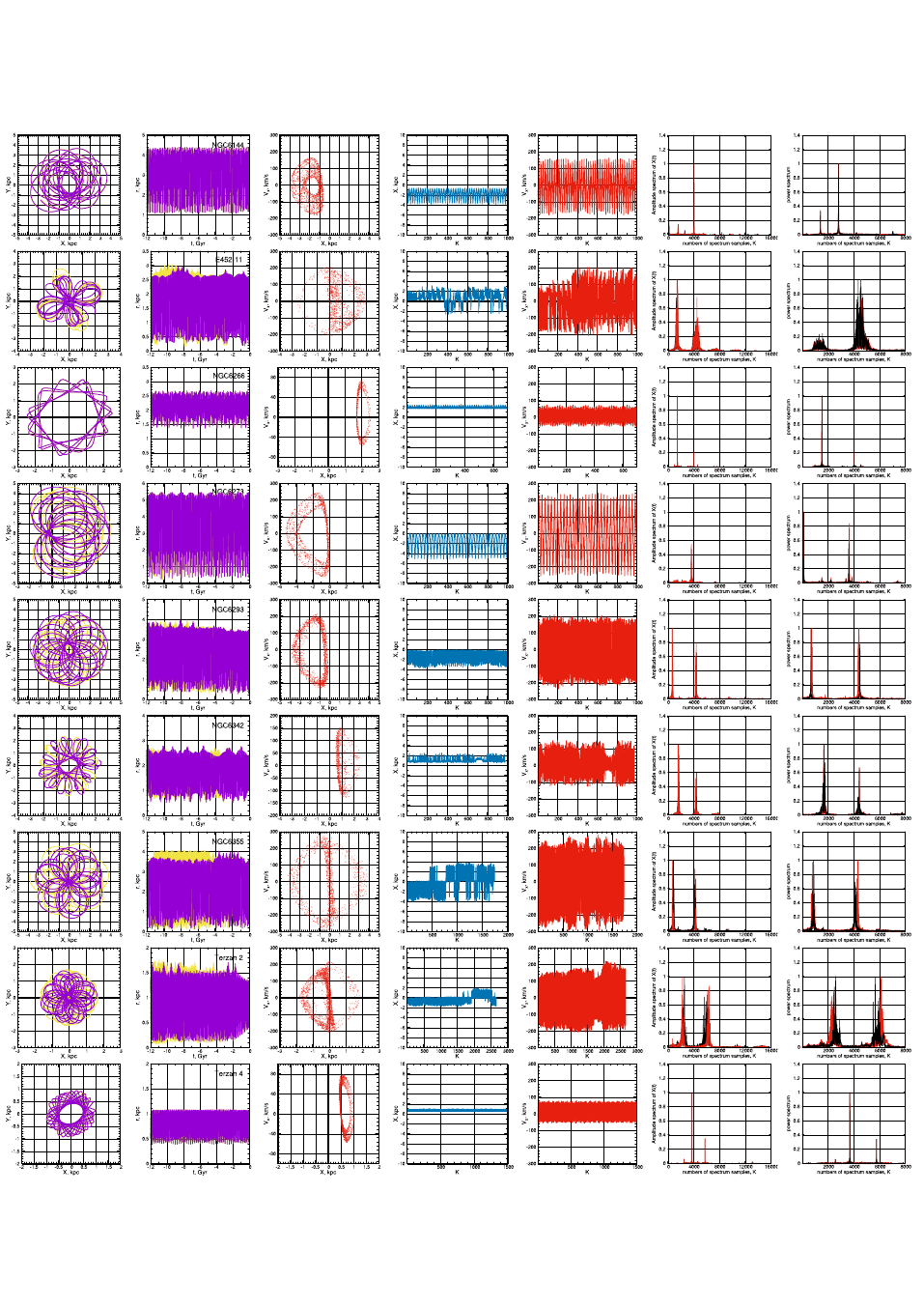}
\vskip -2.55cm
\caption{\scriptsize  Orbits of globular clusters. In the panels from left to right: column (1), projections of orbits onto the plane ($X-Y$ ); column (2) radial values of the initial (reference) and perturbed (shadow) orbits as a function of time (reference orbits are shown in yellow, shadow orbits in purple); column (3), Poincar\'e sections $X-V_x$; column (4), $X$ - coordinates of Poincar\'e sections; column (5), $V_x$ - coordinates of the Poincar\'e sections; column (6), normalized power spectra - the values of the reference and shadow orbits as functions of time, shown in black and red, respectively; and column (7), illustration of the frequency method
(the power spectrum of the first half of the time sequence is shown in red and the second half is shown in black). The names of
the GCs are shown in the second panels from the left.
}
 \label{fD}
\end{center}}
\end{figure*}

\newpage
    \voffset=-4.40truecm

\begin{figure*}
{\begin{center}
     \includegraphics[width=1.0\textwidth,angle=0]{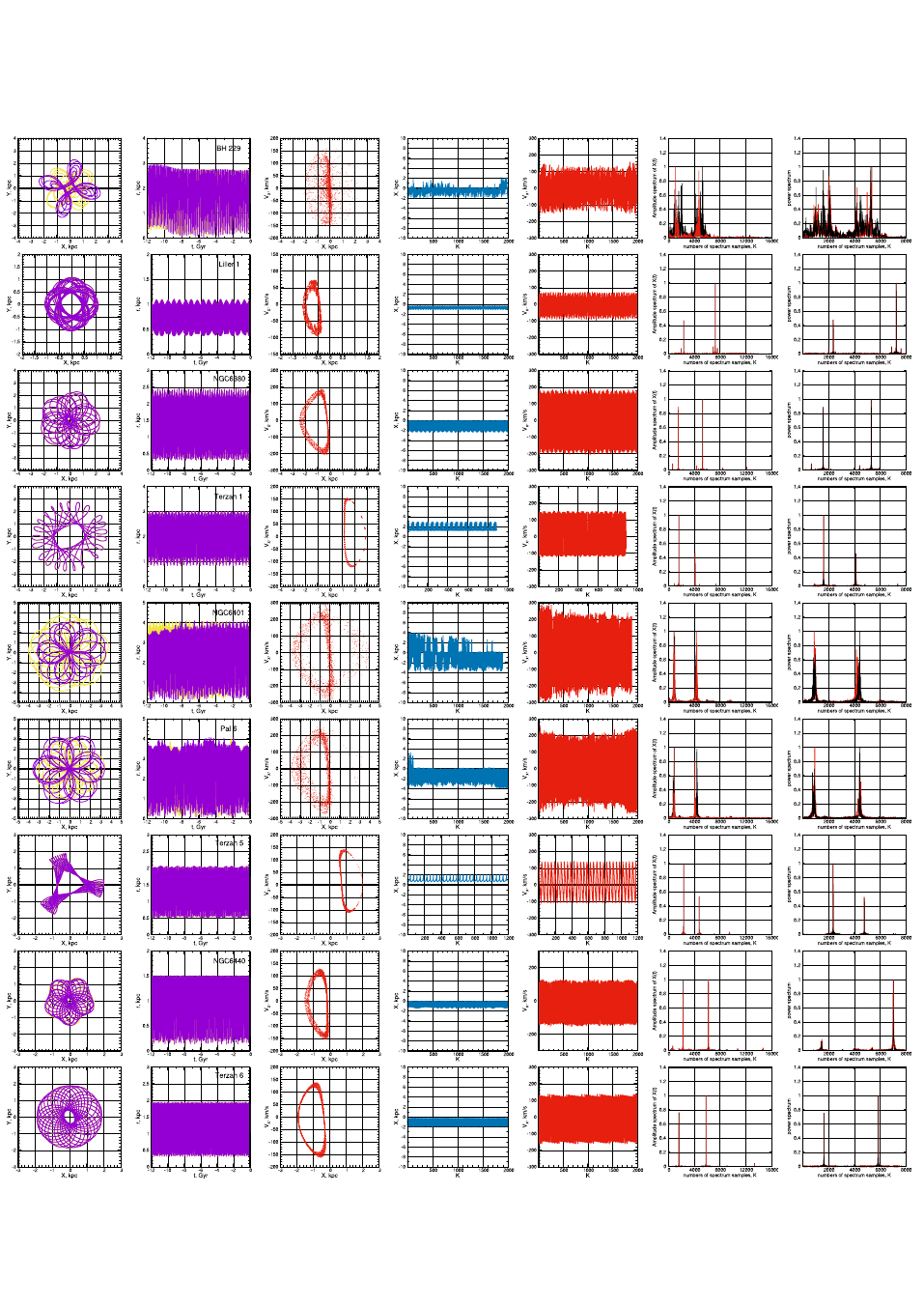}
\vskip -2cm
\centerline{\small Figure 3: (Contd.)}
 \label{fD}
\end{center}}
\end{figure*}

\newpage
    \voffset=-4.40truecm

\begin{figure*}
{\begin{center}
     \includegraphics[width=1.0\textwidth,angle=0]{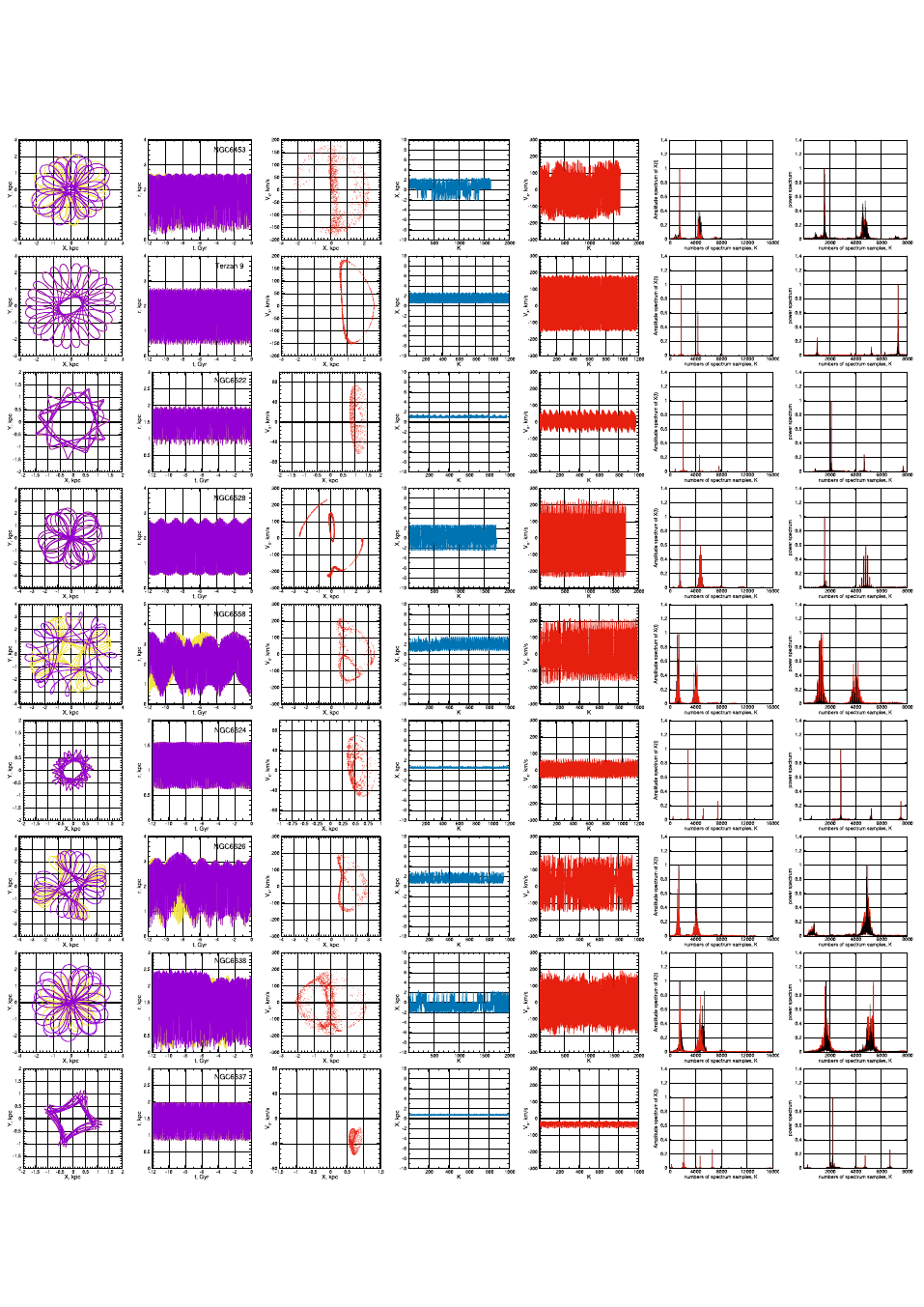}
\vskip -2cm
\centerline{\small Figure 3: (Contd.)}
 \label{fD}
\end{center}}
\end{figure*}

\newpage
    \voffset=-4.40truecm

\begin{figure*}
{\begin{center}
     \includegraphics[width=1.0\textwidth,angle=0]{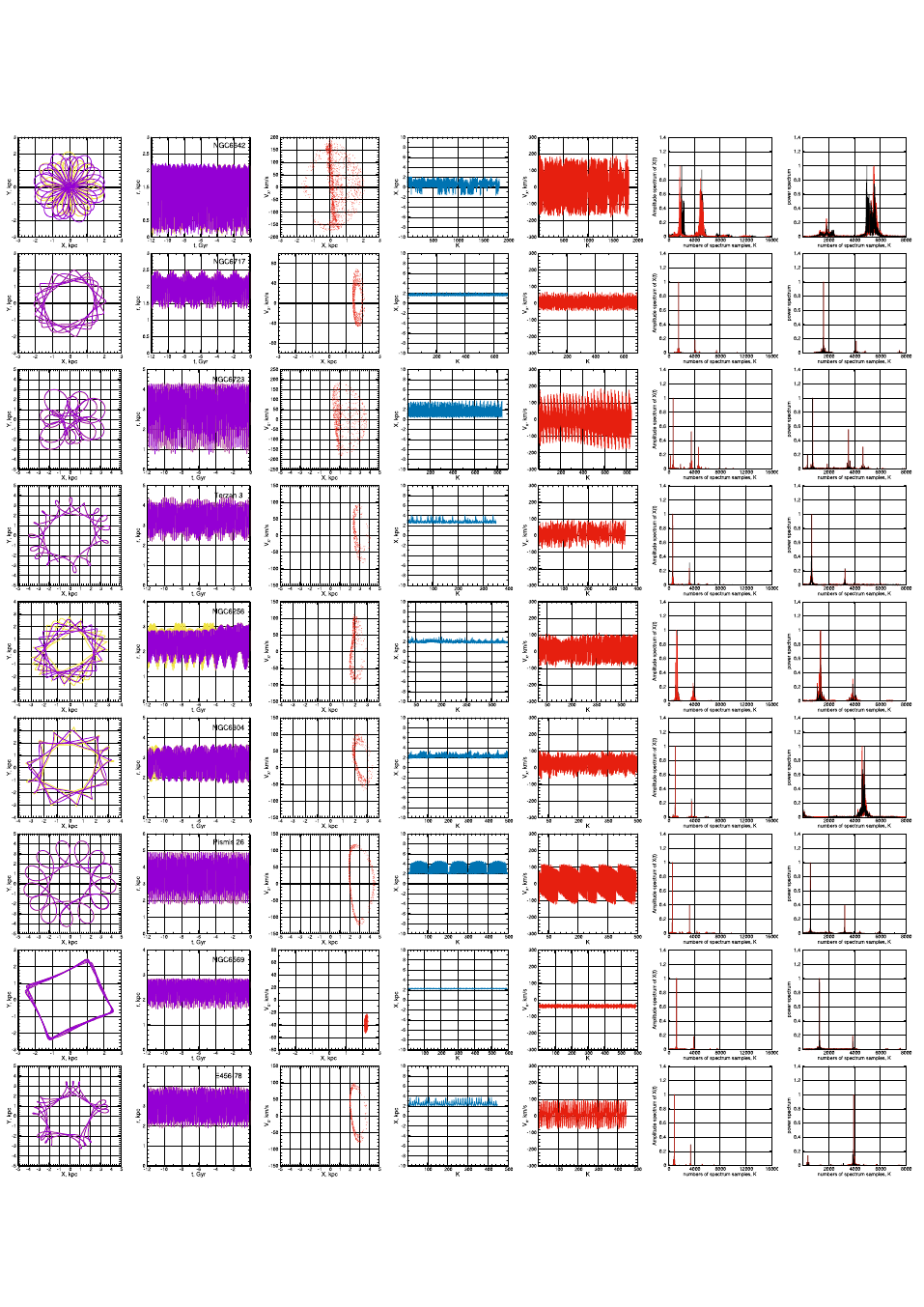}
\vskip -2cm
\centerline{\small Figure 3: (Contd.)}
 \label{fD}
\end{center}}
\end{figure*}

\newpage
    \voffset=-4.40truecm

\begin{figure*}
{\begin{center}
     \includegraphics[width=1.05\textwidth,angle=0]{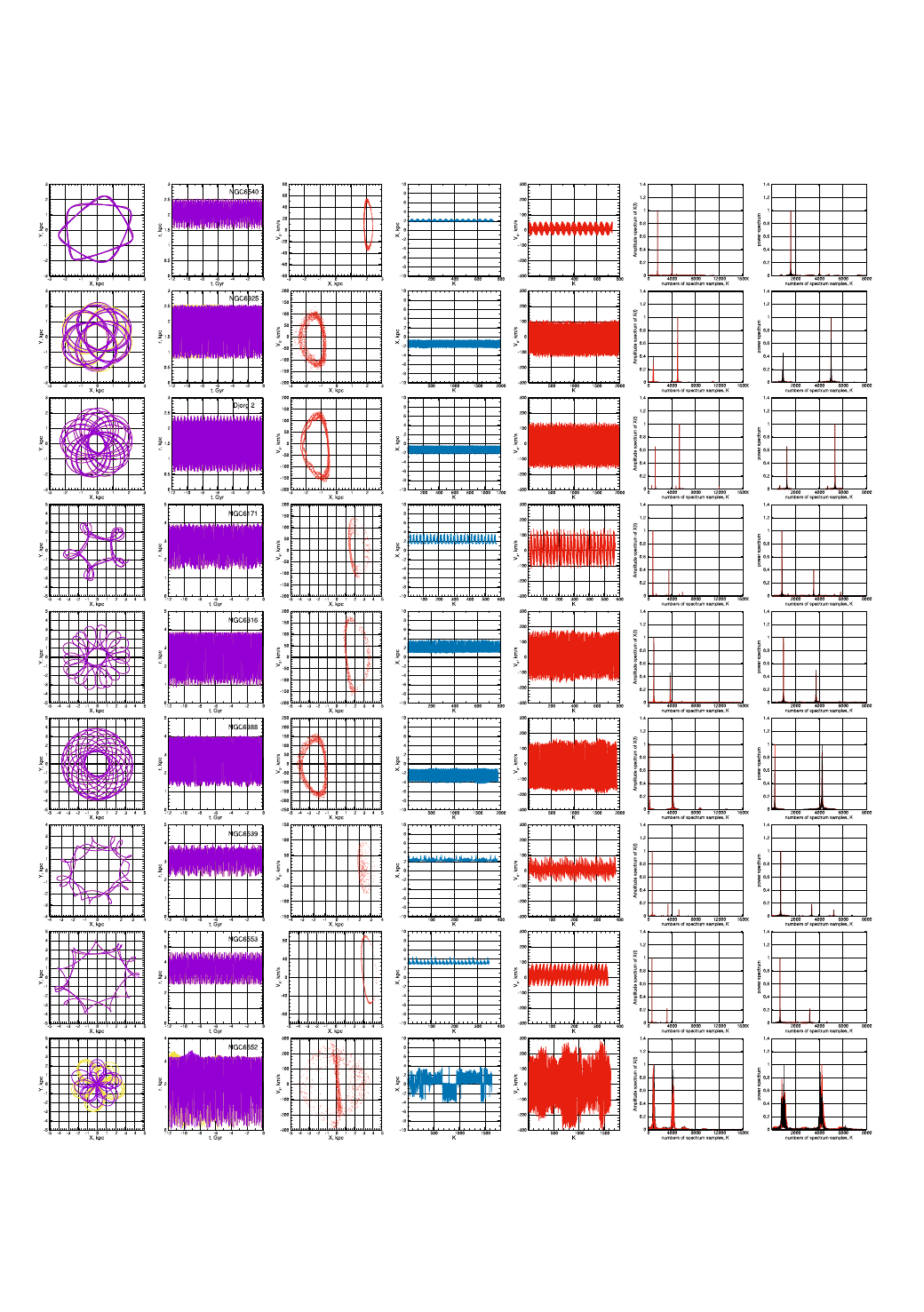}
\vskip -3cm
\centerline{\small Figure 3: (Contd.)}
 \label{fD}
\end{center}}
\end{figure*}

\newpage

\begin{figure*}
{\begin{center}
     \includegraphics[width=0.75\textwidth,angle=-90]{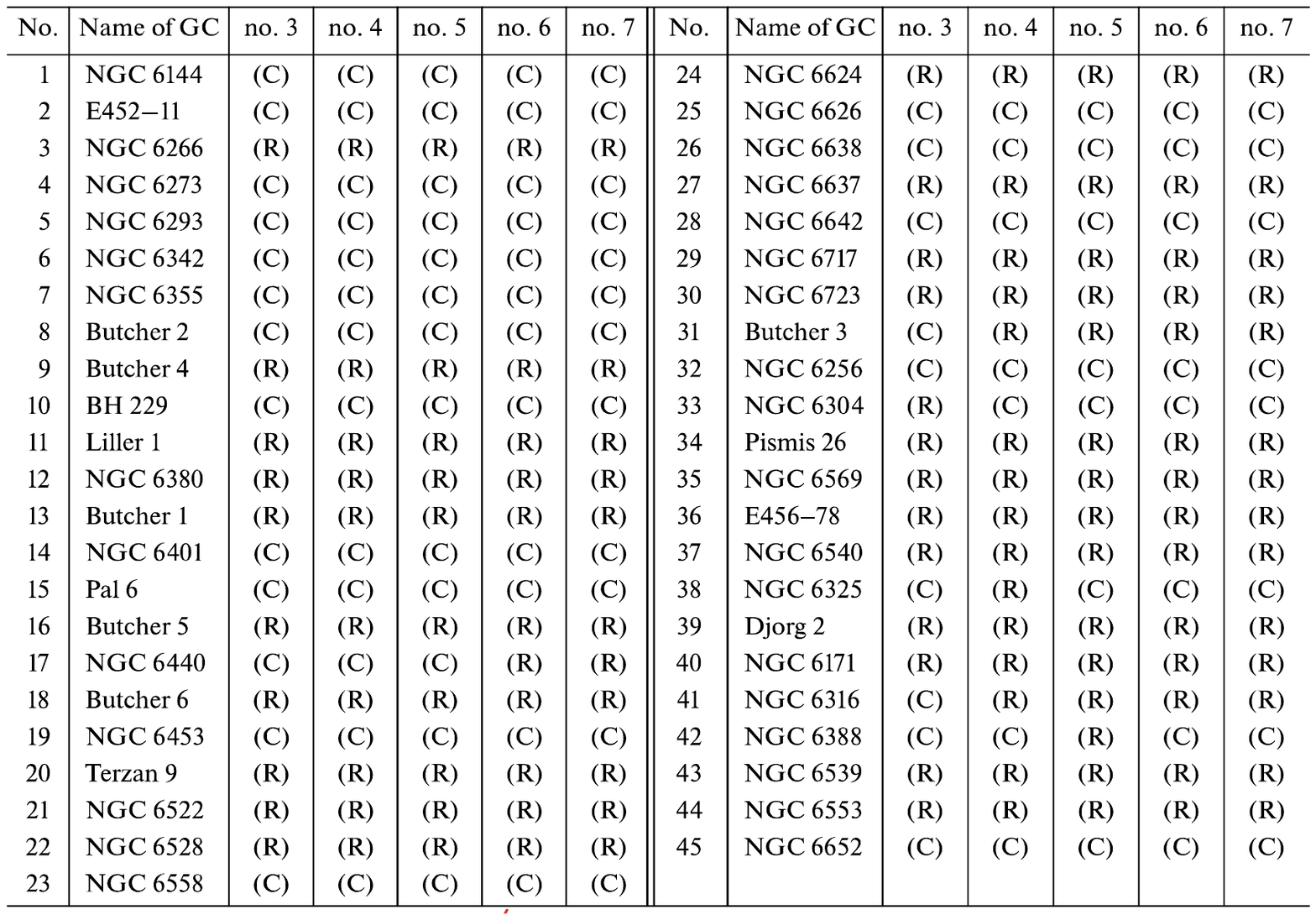}

\vskip 1cm

{\small {\bf Table 2.} Comparative table of the signs of regularity (R) and chaos (C) of the orbits of 45 GCs in the central region of the Galaxy that are obtained by various methods. Column 3 shows the estimates for entropy (8) of spectrum (7); in column 4, the estimate for frequency drift [1]; in column 5, the visual estimate [1]; in column 6, the estimate based on the Poincar\'e section [1]; in column 7, the estimate based on spectra (9) and (10).}
\end{center}}
\end{figure*}
\end{document}